\documentclass[11pt]{article}

\def\tr{\;{\rm tr}\;}

\def\bra{\langle}   \def\ket{\rangle}

\newcommand{\beq}{\begin{eqnarray}}
\newcommand{\eeq}{\end{eqnarray}}

\newcommand{\dd}[2]{\frac {\partial #1}{\partial #2}}
\newcommand{\pdr}{\partial}

\newcommand{\half}{\frac{1}{2}}

\newcommand{\ov}[1]{\frac{1}{#1}}
\newcommand{\fr}[2]{\frac{#1}{#2}}

\def\a{\alpha}      \def\b{\beta}   \def\g{\gamma}       \def\G{\Gamma}
\def\d{\delta}      \def\D{\Delta}  \def\eps{\epsilon} 
                \def\La{\Lambda}
                      
\def\n{\nu}

\def\pr{\prime}

\newcommand{\Ntr}{{{1\over N}{\rm tr}}}

\begin{document}
\input{epsf}

\begin{titlepage}

\title{\normalsize \hfill ITP-UU-06/02  \\ \hfill SPIN-06/02
\\ \hfill {\tt hep-th/0601216}\\ \vskip 0mm \Large\bf
Phase transition in matrix model with logarithmic action: Toy-model
for gluons in baryons}

\author{Govind S. Krishnaswami}
\date{\normalsize Institute for Theoretical Physics \& Spinoza Institute \\
Utrecht University, Postbus 80.195, 3508 TD, Utrecht, The
Netherlands
\smallskip \\ e-mail: \tt g.s.krishnaswami@phys.uu.nl \\ January 29, 2006}

\maketitle

\begin{quotation} \noindent {\large\bf Abstract } \medskip \\

We study the competing effects of gluon self-coupling and their
interactions with quarks in a baryon, using the very simple setting
of a hermitian $1$-matrix model with action $\tr A^4 - \log \det(\nu
+ A^2)$. The logarithmic term comes from integrating out $N$ quarks.
The model is a caricature of 2d QCD coupled to adjoint scalars,
which are the transversely polarized gluons in a dimensional
reduction. $\nu$ is a dimensionless ratio of quark mass to coupling
constant. The model interpolates between gluons in the vacuum
$(\nu=\infty)$, gluons weakly coupled to heavy quarks (large $\nu$)
and strongly coupled to light quarks in a baryon ($\nu \to 0$). It's
solution in the large-$N$ limit exhibits a phase transition from a
weakly coupled $1$-cut phase to a strongly coupled $2$-cut phase as
$\nu$ is decreased below $\nu_c = 0.27$. Free energy and correlation
functions are discontinuous in their third and second derivatives at
$\nu_c$. The transition to a two-cut phase forces eigenvalues of $A$
away from zero, making glue-ring correlations grow as $\n$ is
decreased. In particular, they are enhanced in a baryon compared to
the vacuum. This investigation is motivated by a desire to
understand why half the proton's momentum is contributed by gluons.

\end{quotation}

\vfill \flushleft




\thispagestyle{empty}

\end{titlepage}

\eject

\section{Introduction}

Photons carry a negligible amount of momentum in a moving atom. By
contrast, it is experimentally found \cite{cteq-handbook} that
gluons carry about half the momentum of a nucleon when observed at
momentum transfers of order $Q^2 \sim 1 ~(GeV)^2.$ The growth of the
gluon momentum contribution as $Q^2$ is increased is correctly
predicted by perturbative QCD. However, the momentum fraction
($x_{\rm Bj}$) dependence of gluons in a nucleon at any fixed value
of $Q^2$ appears to be essentially non-perturbative. More generally,
determining the `emergent' bound-state structure of gluons in
baryons from QCD remains an interesting and challenging open problem
of theoretical physics. This problem in $3$ or $4$ dimensional QCD
is still very hard to address analytically. However, there is a
context where it can at least be finitely formulated. This is in the
class of theories where 2d QCD is coupled to adjoint scalar fields.
The principal examples are the reduction of QCD from $3$ or $4$ to
$2$ dimensions, where we assume that fields are independent of
transverse coordinates \footnote{This dimensional reduction may well
provide a first approximation to the gluon distribution of the
proton measured in the Bjorken limit of deep inelastic scattering.}.
In this case, the longitudinal component of the gauge field
contributes a linear potential between dynamical quarks $q^a(x)$ and
the transverse components transform as adjoint scalar fields
$A^a_b(x)$, for instance in the light-cone gauge. Thus, for $N$
colors $a,b$, we have a two-dimensional field theory of $N\times N$
hermitian matrix-valued and $N$-component vector-valued fields with
only a `global' $U(N)$ invariance. The gauge-invariant observables
include glue-ring (closed-string) variables
    \beq
        tr [A(x_1) A(x_2) \cdots A(x_n)] \label{e-glue-ring}
    \eeq
and meson (open-string) variables
    \beq
     ~q^\dag_{a_1}(x_0) ~A^{a_1}_{a_2}(x_1)~ A^{a_2}_{a_3}(x_2)
        \cdots A^{a_n}_{a_{n+1}}(x_n) ~q^{a_{n+1}}(x_{n+1})
        \label{e-meson}
    \eeq
where the points $x_i$ are null separated\footnote{The parallel
transport operators between these points are trivial in a gauge
where the null component of the the gauge field is set to zero.}. In
't Hooft's large-$N$ limit, the expectation values of products of
these variables factorize. So we may restrict attention to the
single trace observables. Quarks are suppressed by one power of $N$
in calculating vacuum expectation values of glue-ring observables in
the large-$N$ limit \cite{tHooft-large-N}. However, when the vacuum
is replaced by a baryon state, interactions with quarks are just as
important as gauge field self-interactions
\cite{Witten-baryons-largeN}. For definiteness, let us consider 2d
QCD coupled to a single adjoint scalar field via the action
(non-dynamical fields have been eliminated)
    \beq
    S = \int dt~ dx \tr \bigg[q^\dag i \partial_t q + \half m^2
        q^\dag \ov{i \pdr_x} q + \partial_x A \partial_t A \bigg]
        - \fr{g^2}{2} \int dt dx dy ~J^a_b(x) ~{|x-y| \over 2} ~J^b_a(y)
        \label{e-adjoint-qcd-lagrangian}
    \eeq
where the current
    \beq
    J^a_b(x) = i [A(x),\partial_x A(x)]^a_b - q^\dag_b(x) q^a(x).
    \eeq
$m$ is the `current' quark mass and $g$ is a coupling constant with
the dimensions of mass. There is no term of the form $q^\dag A q$
when non-dynamical fields are eliminated. This is essentially the
dimensional reduction from $3$ dimensions, except for the absence of
terms of the form $q^\dag A \ov{\pdr_x} A q$ and $q^\dag A
\ov{\pdr_x} q$. These theories have no ultraviolet divergences,
though $m$ undergoes a finite renormalization ($m^2 \to m^2 - g^2
N/\pi$). Heuristically, $g^2$ is related to the dimensionless 4d
coupling constant $g^2_4$ via a factor of the transverse area of
hadrons, $g^2 \sim \La_{\rm QCD}^{2} g^2_4$.

Before proceeding further, let us mention some literature on these
theories. The adjoint scalar glueball spectrum was studied by
numerical diagonalization of the hamiltonian by Klebanov and Dalley
\cite{klebanov-dalley}. There is a large literature on discretized
light-cone and transverse lattice approaches for which we refer to
the review by Burkardt and Dalley \cite{burkardt-dalley}. Among the
more analytical approaches, Rajeev, Turgut and Lee
\cite{rajeev-turgut-lee} studied the Poisson and Lie algebras of
Wilson loops, open and closed string observables in the large-$N$
limit of the hamiltonian framework. On the other hand, Rajeev
formulated $2$ dimensional QCD in the large-$N$ limit (without the
adjoint scalars) as a non-linear classical theory of gauge-invariant
quark bilinears \cite{rajeev-2d-qhd}. By developing approximation
methods to solve this non-linear classical theory it was possible to
analytically determine the quark structure of the baryon predicted
by $2$d QCD. This was shown to provide a first approximation to the
non-perturbative quark structure functions of baryons measured in
deep inelastic scattering \cite{soliton-parton,gsk-thesis}.

Despite these developments, determining the expectation values of
glue-ring and meson observables analytically remains quite hard for
two reasons. On the one hand, there are an infinite number of
observables with ever increasing string length. Their dynamics is
inter-linked via the factorized Schwinger-Dyson or loop equations.
On the other hand, we are interested not just in the vacuum
correlations but those in the ground state of the baryon, which has
itself to be determined dynamically.

We propose to use the approximate ground-state of the baryon $|\Psi
\ket$ determined in the large-$N$ limit of $2$d QCD as a starting
point for the harder problem of determining the gluon correlations
in the baryon. We have shown \cite{soliton-parton,gsk-thesis} that
    \beq
    |\Psi \ket = \int dx_1 \cdots dx_N \eps^{a_1 \cdots a_N}
        \psi(x_1, \cdots x_N) q_{a_1}^* \cdots q_{a_N}^*(x_N)
    \eeq
for a factorized wave function (see
Sec.~\ref{s-ansatz-for-baryon-state}) provides a good approximation
to the ground state of the baryon in the chiral and large-$N$
limits. In a sense, the state $|\Psi \ket$ contains no gluons.
Nevertheless, the presence of a baryon $|\Psi \ket$ made of $N$
quarks may be expected to deform the vacuum of the gluon field. The
idea is to approximately determine the large-$N$ expectation values
of glue-ring observables (\ref{e-glue-ring}) in the background of
this $N$ quark baryon. Though $|\Psi \ket$ is not the true baryon
ground state of the action (\ref{e-adjoint-qcd-lagrangian}), it is
still an approximate ground state of the baryon number one sector of
2d QCD in the large-$N$ limit. As such, it furnishes a variational
approximation to the ground state of
(\ref{e-adjoint-qcd-lagrangian}). In this manner, we hope to isolate
a portion of this difficult problem that we have some chance of
addressing analytically.

In this paper we will take a very modest step towards the
above-mentioned goal by studying a much simplified version via the
path integral approach in the large-$N$ limit. To focus on the
interesting baryon-part of the problem, we study a caricature of the
action (\ref{e-adjoint-qcd-lagrangian}) without derivatives or
non-local interactions and where fields are assumed time
independent:
    \beq
    S(A,q,q^*)  = \tr \int dx \bigg[ \a A^4(x)
        - i q^*(x) (m + \a A^2(x)) q(x)\bigg].
    \eeq
Here $\a$ is the 't Hooft coupling with dimensions of mass, held
fixed as $N \to \infty$. Though an oversimplification, it allows us
to isolate the dynamics associated to the degrees of freedom at the
locations of the $N$ quarks making up the baryon. After integrating
out the quarks we get an $N$ matrix model with an action involving
not just traces of powers of the adjoint scalar, but also its
inverse and determinant. To get a computable toy-model that
preserves the essential picture of gluon correlations in a baryon
background, we assume that the adjoint scalar field is equal at the
positions of the different quarks. This leads us to a $1$-matrix
model whose action is a quartic polynomial with a logarithmic
modification.
    \beq
        S(A) = \tr\bigg[A^4 - \log{[\nu + A^2]}\bigg]
    \eeq
where the dimensionless parameter $\nu$ is a ratio of quark mass to
coupling constant. The logarithmic term encapsulates the effect of
the $N$-quark baryon. Due to the significant truncations we make,
this matrix model is not an approximation to the $1+1$ dimensional
field theory. It is the simplest toy-model where we may study the
competing effects of gluon self-interactions and their interaction
with quarks in a baryon. Finite-dimensional matrix models are often
of independent interest. For instance, a slightly different
logarithmic action $\log(1 - A) + A$ appears in the Penner matrix
model, which found applications to $c=1$ string theory
\cite{penner-distler-vafa}.

{\em Summary of Results:} We solve our $1$-matrix model in the
large-$N$ limit by an extension of the methods used by Brezin et.
al. \cite{bipz} for polynomial potentials. This is possible because
the derivative of the action is a rational function. We determine
the free energy and glue-ring expectation values (moments of the
adjoint scalar) in this caricature of a baryon. As $\nu$ decreases
below $\n_c = 0.27$, the 1-cut solution\footnote{One and two `cuts'
refer to the number of disjoint intervals of the real line on which
the distribution of eigenvalues of $A$ is supported.} of the matrix
model makes a transition to a 2-cut solution. The reason is that the
logarithmic term in the action (due to the presence of the baryon)
makes small eigenvalues for the adjoint scalar energetically costly.
The 1-cut solution could have been obtained by summing planar
diagrams, but the 2-cut solution is analytically unrelated and could
not be obtained that way. At the critical point $\n_c$, the leading
discontinuity in free energy is in its $3^{\rm rd}$ derivative,
while the $2^{\rm nd}$ derivatives of the two- and four-point
correlations are discontinuous. This large-$N$ phase transition
bears some resemblance to the Gross-Witten transition in the unitary
one-plaquette model \cite{gross-witten} or that in the hermitian
$m^2A^2 + g A^4$ matrix model as $m^2/\sqrt{g}$ is made sufficiently
negative \cite{cicuta-et-al}. All these phase transitions involve a
jump discontinuity in the third derivative of free energy. One
consequence is that analytic continuation from a weak-coupling
treatment of gluons, starting from the vacuum ($\n = \infty$) would
not be successful in predicting their behavior when strongly coupled
to light quarks in a baryon ($\n \to 0^+$). Moreover, moments of the
adjoint scalar are enhanced in the baryon compared to the vacuum,
since, in the two-cut phase, the support of the eigenvalue
distribution excludes the origin.

{\em Organization of article:} We begin with our ansatz for the
baryon state in Sec.~\ref{s-ansatz-for-baryon-state} and give a path
integral formulation of the problem of finding glue-ring
correlations in Sec.~\ref{s-gluon-corrln-in-baryon-state}.
Sec.~\ref{s-mat-model-caricature} describes the truncations and
approximations that lead us to a matrix model for gluon correlations
in a baryon. Sections \ref{s-1-cut-soln} and \ref{s-2-cut-soln} give
the solution of the matrix model in the weak-coupling ($\n \geq
\n_c$) and strong-coupling $(\n \leq \n_c)$ phases respectively. The
heavy quark limit, neighborhood of the critical point, and the
chiral limit are treated in Sec.~\ref{s-three-spl-cases}.
Sec.~\ref{s-discussion} presents a discussion of our results and
open questions.

\section{Ansatz for baryon state}
\label{s-ansatz-for-baryon-state}

A lesson from our study of 2d QCD in the large-$N$ limit is that the
ground state of the baryon is well approximated by a state
containing $N$ `valence' quarks. Such a state may be built out of
    \beq
        |\stackrel{x_1}{a_1}; \cdots; \stackrel{x_N}{a_N} \rangle ~=~
        \hat q^\dag_{a_1}(x_1) \hat q^\dag_{a_2}(x_2) \cdots \hat
        q^\dag_{a_N}(x_N)~ |0 \ket
    \eeq
where $\hat q, \hat q^\dag$ satisfy canonical anti-commutation
relations with respect to the Dirac vacuum $|0 \ket$. Ignoring
flavor and spin quantum numbers, a general $N$-quark state is a
linear combination
    \beq
        |\Psi \ket = \hat \Psi |0 \ket = \int dx_1 \cdots dx_N
        \Psi(\stackrel{x_1}{a_1};
        \cdots; \stackrel{x_N}{a_N}) \hat q^\dag_{a_1}(x_1) \hat q^\dag_{a_2}(x_2)
        \cdots \hat q^\dag_{a_N}(x_N)~ |0 \ket.
    \eeq
To represent a baryon, $\Psi$ must be totally anti-symmetric in
color
    \beq
        \Psi(\stackrel{x_1}{a_1}; \cdots; \stackrel{x_N}{a_N}) =
        \eps^{a_1 \cdots a_N} \psi(x_1, \cdots, x_N),
    \eeq
and therefore represents a fermion if $\psi(x_1, \cdots, x_N)$ is a
symmetric function. For a path integral formulation, associate the
grassmann-valued field $q^a(x)$ and its complex conjugate $q^*_a(x)$
to the operators $\hat q^a(x)$ and $\hat q^\dag_a(x)$. Then the
baryon state $|\Psi \ket$ becomes
    \beq
        |\Psi \ket ~ \mapsto ~ \int dx_1 \cdots dx_N ~ \eps^{a_1 \cdots
        a_N} ~ \psi(x_1 \cdots x_N) ~ q_{a_1}^*(x_1) q_{a_2}^*(x_2)
        \cdots q_{a_N}^*(x_N).
    \label{e-baryon-state}
    \eeq
By solving 2d QCD in the large-$N$ limit we determined that
$\psi(x_1 \cdots x_N)$ is well-approximated by an $N$-fold product
of single particle wave-functions $\psi(x)$. A way to understand
this is that once the antisymmetry in color is accounted for, quarks
behave like $N$ bosons which condense to the same one-particle
ground-state. A good variational estimate turns out to be $\psi(x) =
\ov{\sqrt{\pi}} \ov{(1-i x)}$ which is the Fourier transform of the
`valence quark wave-function' $\tilde \psi(p) = 2 \sqrt{\pi} e^{-p}~
\theta(p \geq 0)$. Though presented in the language of quantum
many-body theory, the above ansatz may be obtained as the first term
in a systematic field theoretic approximation method for the ground
state of 2d QCD (see \cite{gsk-thesis} for details). One goal is to
have a similar understanding of the gluon content of the baryon from
the adjoint scalar field theory (\ref{e-adjoint-qcd-lagrangian}),
which may then also be interpreted in terms of a relativistic
many-body problem.

\section{Gluon correlations in the baryon state}
\label{s-gluon-corrln-in-baryon-state}

For any $U(N)$-invariant operator $\mathcal{O}$, such as the
glue-ring (\ref{e-glue-ring}), the relation between expectation
values of Heisenberg field operators and functional integrals in
Euclidean space-time is
    \beq
        \bra \xi| \mathcal{O}(A,q,q^*) |\xi \ket &=& \bra0| \hat \xi \mathcal{O} \hat \xi^\dag
            |0 \ket = \ov{Z} \int \mathcal{D}A \mathcal{D}q \mathcal{D}q^*
            e^{-N S(A,q,q^*)} \xi \mathcal{O}(A,q,q^*) \xi^* \cr
        {\rm where~~} Z &=& \int \mathcal{D}A \mathcal{D}q \mathcal{D}q^*
            e^{-S(A,q,q^*)} \xi \xi^*
    \eeq
Here $\xi^\dag$ creates a baryon in the infinite past $-T \to
-\infty$ while $\xi$ annihilates it in the infinite future $T$.
Using periodic boundary conditions in time we can assume that these
operators are adjoints of each other and evaluated at a common time
$T$. Specializing to our baryon state $|\xi \ket = |\Psi \ket$ given
in (\ref{e-baryon-state}),
    \beq
         \bra \Psi| \mathcal{O}(A,q,q^*) |\Psi \ket &=& \ov{Z} \int \mathcal{D}A \mathcal{D}q \mathcal{D}q^*
            e^{-N S(A,q,q^*)} \int dx_1 \cdots dx_N \eps_{a_1 \cdots
            a_N} \psi(x_1,\cdots,x_N) \cr && \times q^{a_1}(x_1,T) \cdots
            q^{a_N}(x_N,T) ~ \mathcal{O}(A,q,q^*) \int dy_1 \cdots dy_N \eps^{b_1 \cdots
            b_N} \cr && \times \psi^*(y_1,\cdots,y_N) q_{b_1}^*(y_1,T) \cdots
            q_{b_N}^*(y_N,T)
    \eeq
The functional integrals over Grassmann variables would vanish if
$x_i \ne y_i$, so we take them to be equal and get
    \beq
        \bra \Psi| \mathcal{O}(A,q,q^*) |\Psi \ket &=& \ov{Z} \int dx_1 \cdots
        dx_N |\psi(x_1, \cdots x_N)|^2 \int \mathcal{D}A \mathcal{D}q \mathcal{D}q^*
        e^{-N S(A,q,q^*)} ~ \eps_{a_1 \cdots a_N} \eps^{b_1 \cdots b_N}
        \cr && \times ~
        q^{a_1}(x_1,T) \cdots q^{a_N}(x_N,T) ~ \mathcal{O}(A,q,q^*) ~
        q_{b_1}^*(x_1,T) \cdots q_{b_N}^*(x_N,T)
    \eeq
where
    \beq
        Z &=& \int dx_1 \cdots
        dx_N |\psi(x_1, \cdots x_N)|^2 \int \mathcal{D}A \mathcal{D}q \mathcal{D}q^*
        e^{- N S(A,q,q^*)} ~ \eps_{a_1 \cdots a_N} ~ \eps^{b_1 \cdots b_N}
        \cr && \times ~
        q^{a_1}(x_1,T) \cdots q^{a_N}(x_N,T) ~
            q_{b_1}^*(x_1,T) \cdots q_{b_N}^*(x_N,T) \cr
    &=& \int dx_1 \cdots dx_N ~|\psi(x_1, \cdots x_N)|^2 ~Z(x_1,\cdots, x_N)
    \eeq
Now, we set aside the integral over $x_1, \cdots, x_N$ and focus on
the functional integral\footnote{One complication that we do not
address in this paper is that in calculating expectation values, the
functional integral is not separately normalized, but only after
integration over the points $x_1, \cdots, x_N$.}
    \beq
    Z(x_1, \cdots x_N) &=& \int \mathcal{D}[A q q^*]
        e^{-N S(A,q,q^*)}  \eps_{a_1 \cdots a_N} ~ \eps^{b_1 \cdots b_N}
        \cr && q^{a_1}(x_1,T) \cdots q^{a_N}(x_N,T)
        q_{b_1}^*(x_1,T) \cdots q_{b_N}^*(x_N,T)
    \label{e-Z-of-x1-thro-xN}
    \eeq

\section{Matrix model caricature of baryon}
\label{s-mat-model-caricature}

Physically, we are primarily interested in the shape and energy of
the proton in its stationary ground state, which is probably best
addressed in a hamiltonian framework. To mimic this in our
path-integral approach, we will simply assume that fields are
independent of time and ignore the $\pdr_t$ terms in
(\ref{e-adjoint-qcd-lagrangian}), which contain the symplectic
structure in a hamiltonian approach. This can be regarded either as
a crude passage to a kind of hamiltonian or as a further dimensional
reduction. Moreover, it is clear that the locations of the quarks
$x_1, \cdots, x_N$ play a special role in the path integral
(\ref{e-Z-of-x1-thro-xN}). We would like to isolate the contribution
of those points. Of course, this is not possible with the Lagrangian
(\ref{e-adjoint-qcd-lagrangian}), since it involves spatial
derivatives as well as non-local interactions. However, in keeping
with the modest goals of this paper, we work with an `action' that
is motivated by (\ref{e-adjoint-qcd-lagrangian}), but does not
involve any derivatives or non-local interactions
    \beq
    S(A,q,q^*) = \tr \int dx \bigg\{V(A)
        - i q^* (m + \a A^2) q  \bigg\}
    \eeq
In this caricature, $V(A) = \a A^4$ stands in place of the
current-current self-interaction of the adjoint scalars in
(\ref{e-adjoint-qcd-lagrangian}). $q^* A^2 q$ models the quark-gluon
cross term in the current-current interaction. We ignore the quartic
term in quarks, which does not involve gluons. The primary effect of
this term has already been taken into account in determining the
baryon state $|\Psi \ket$ (\ref{e-baryon-state}) by solving 2d QCD
in the large-$N$ limit \cite{rajeev-2d-qhd}. Thus
(\ref{e-Z-of-x1-thro-xN}) becomes
    \beq
    Z(x_1,\cdots,x_N) &=& \int \mathcal{D}A \mathcal{D}q^* \mathcal{D}q~
        e^{-N\int dx[\tr V(A) - iq^* (m + \a A^2) q]} \cr
        && \times \eps^{a_1 \cdots a_N} \eps_{b_1 \cdots b_N}
        q^{b_N}(x_N) \cdots q^{b_1}(x_1) q_{a_1}^*(x_1) \cdots
        q_{a_N}^*(x_N)
    \eeq
The partition function can be written as:
    \beq
         Z(x_1,\cdots,x_N) = \int \mathcal{D}A e^{-N \tr \int dx V(A)} Z_q
    \eeq
where $Z_q$ is the `quark part' of the partition function
    \beq
        Z_q &=& \int \mathcal{D}q^* \mathcal{D}q e^{iN \int dx
        q^*(m+ \a A^2)q} \eps^{a_1 \cdots a_N} \eps_{b_1 \cdots b_N}
         q^{b_N}(x_N) \cdots q^{b_1}(x_1) q_{a_1}^*(x_1) \cdots
            q_{a_N}^*(x_N) \cr
        &=& \prod_{x} \int dq^*(x) dq(x) e^{iN \D q^*(m+  \a A^2)q
        } \eps^{a_1 \cdots a_N} \eps_{b_1 \cdots b_N}
        \cr && q^{b_N}(x_N) \cdots q^{b_1}(x_1) q_{a_1}^*(x_1) \cdots
            q_{a_N}^*(x_N).
    \eeq
In the second line we have discretized $\int dx$ to a sum $\sum_{x}
\D$, where $\D \sim dx$. Moreover, $Z_q$ may be factored into a
`vacuum part' and a `baryon part' $Z_q = Z_{qv} Z_{qb}$.
    \beq
    Z_{qv} &=& \prod_{x\ne x_i} \int dq^*(x) dq(x)
                e^{i N \D q^*(m + \a A^2)q}
           = \prod_{x\ne x_i} \det{[N\D(m + \a A^2(x))]} \cr
    Z_{qb} &=& \eps^{a_1 \cdots a_N} \eps_{b_1 \cdots b_N}
            \prod_{l=1}^N \int dq^*(x_l) dq(x_l)
        e^{i N \D q^*(m + \a A^2)q}
        q^{b_l}(x_l) q_{a_l}^*(x_l).
    \eeq
Now we use the result $\int dq^* dq e^{i q^* B q} q^b q_a^* =
{[B^{-1}]}_a^b~\det{[B]}$ to get
    \beq
    {Z_{qb} \over (N\D)^{N^2-N}} = \eps^{a_1 \cdots a_N} \eps_{b_1 \cdots b_N}
                \det{B(x_1)} {[B^{-1}(x_1)]}_{a_1}^{b_1} \cdots
                \det{B(x_N)} {[B^{-1}(x_N)]}_{a_1}^{b_1}
    \eeq
where $B(x_i) = m + \a A^2(x_i)$ is an $N \times N$ matrix for each
$i = 1, \cdots N$. We can ignore the constant overall factor
involving $\Delta$. Thus $Z_{qb}$ depends on the gluon field at the
location of the quarks. We factor the integral over $A$ into a
baryon contribution and vacuum contribution.
    \beq
    \int \mathcal{D}A e^{-N \tr \int dx V(A)} &=&
    \prod_{x \ne x_i} \int dA(x) e^{-N \D \tr V(A)} \cr
    && \times \int \prod_{i=1}^N  dA(x_i) e^{-N \D \tr \sum_{i=1}^N
    V(A(x_i))}.
    \eeq
Combining the quark and gluon vacuum contributions, we define the
vacuum part of the partition function as
    \beq
    Z_{vac}(x_1,\cdots,x_N) = \prod_{x \ne x_i} \int dA(x) e^{-N \D \tr V(A)}
        \det{[N\D(m + \a A^2(x_i))]}.
    \eeq
$Z_{vac}$ is, for our purposes, the uninteresting part, so we
consider the quotient
    \beq
        {Z(x_1 \cdots x_N) \over Z_{vac}(x_1,\cdots,x_N)} &=&
         \int \prod_{i=1}^N  dA(x_i) e^{-N \tr \sum_{i=1}^N \Delta V(A(x_i))}
         \eps^{a_1 \cdots a_N} \eps_{b_1 \cdots b_N} \cr
        &&  \times \det{B(x_1)} {[B^{-1}(x_1)]}_{a_1}^{b_1} \cdots
               \det{B(x_N)} {[B^{-1}(x_N)]}_{a_1}^{b_1}
    \eeq
Here $B(x_i) = m + \a A^2(x_i)$. This is an $N$-matrix model
involving the $N \times N$ matrices $A(x_i)$, $i = 1, \cdots N$. We
do not yet have a way of approximately solving this multi-matrix
model, so we will make some further assumptions that allow us to
reduce it to a one-matrix model.

\subsection{Reduction to a $1$-matrix model}
\label{s-reduc-to-one-mat-mod}

Suppose we assume that the adjoint scalar gluon field is equal at
the positions of the $N$ quarks, $A(x_i) = A$ for $i = 1, \cdots N$.
Then $B(x_1) = B(x_2) = \cdots = B(x_N) \equiv B$ and we can
simplify $Z_{qb}$:
    \beq
    Z_{qb} = (\det{B})^N \eps_{a_1 \cdots a_N} \eps^{b_1 \cdots b_N}
             [B^{-1}]^{a_1}_{b_1} \cdots [B^{-1}]^{a_N}_{b_N}
           = (\det{B})^{N-1}.
    \eeq
Then,
    \beq
        {Z(x_1 \cdots x_N) \over Z_{vac}(x_1,\cdots,x_N)}
    &=& \int dA e^{-N L \tr V(A)}
        (\det{B})^{N-1} \cr
    &=& \int dA e^{-N \bigg[L \tr V(A) - \log\det[m + \a A^2] \bigg]}
    \eeq
where we have replaced $N-1$ by $N$ in anticipation of the large-$N$
limit. $N \Delta \equiv L$ is a `length' of the baryon, assumed to
have a finite limit as $N \to \infty$ and $\D \to 0$. Recalling that
$V(A)=\a A^4$, the action becomes $S(A) = \tr [L\a A^4 - \log(m+ \a
A^2)]$. Re-scaling $A \to A/(\a L)^{1/4}$ we find there is only one
dimensionless parameter $\n = m\sqrt{L}/\sqrt{\a}$ on which the
observables of this matrix model depend non-trivially. The
appearance of $L$ is an artifact of our truncation and could
probably be avoided. In the field theory
(\ref{e-adjoint-qcd-lagrangian}), the relevant parameter would be
$m/g$. Thus we arrive at the one-matrix model
    \beq
    Z = \int dA e^{-N\tr[A^4 - \log[\nu+A^2]]}.
    \eeq
$\n$ as a dimensionless ratio of quark mass to coupling constant.
The absence of a quadratic term in $A$ may be traced to the absence
of a gluon mass term in (\ref{e-adjoint-qcd-lagrangian}). Since $A$
is hermitian, $A^2$ is positive. Thus, there is no difficulty in
defining $\log[\nu + A^2]$ for $\nu > 0$. The expectation values of
$\tr A^n$ in this matrix model are what remain of the glue-ring
expectation values.

\section{One-cut solution of quartic + log matrix model}
\label{s-1-cut-soln}

The rest of this paper will be devoted to a study of the large-$N$
limit of the one-matrix model with action and free energy
    \beq
        S(A) = \tr \bigg[A^4 - \log(\nu + A^2) \bigg], ~~~
        E(\n) = - \lim_{N \to \infty} \ov{N^2} \log \int dA e^{-N
        S(A)}.
    \eeq
Glue-ring expectation values
    \beq
    G_n(\n) = \lim_{N \to \infty} \bigg(\int dA e^{-N S(A)} \Ntr A^n \bigg)/ \int dA e^{-N S(A)}
    \label{e-def-of-Gn}
    \eeq
are given by moments $G_n = \int \rho(x) x^n dx$ of the eigenvalue
density. $\rho(x)$ must minimize the free energy
    \beq
    E[\rho] = \int S(x) \rho(x) dx - \int dx dy \rho(x) \rho(y) \log|x-y|; ~~~
    E(\n) = \min_{\rho}~ E[\rho]
    \eeq
where $S(x) = x^4 - \log{(\nu + x^2)}$. (a) For $\n$ sufficiently
large, $S(x)$ is convex from below and we expect $\rho$ to be
supported on a single interval. (b) For $\n$ sufficiently small (but
positive), $S(x)$ is shaped like a Mexican hat (see
Fig.~\ref{f-rhocrit}) and we expect $\rho$ to be supported on a pair
of intervals located near the minima of $S(x)$. These two cases are
treated in Sec.~\ref{s-1-cut-soln} and \ref{s-2-cut-soln}. For
$\rho$ supported on a single interval, the Mehta-Dyson linear
integral equation for an extremum of $E[\rho]$ is
    \beq
        S'(x) = 4 x^3 - {2 x \over \nu + x^2} = 2 {\cal P}
            \int_{-2a}^{2a} {\rho(y) \over x-y} dy ~~~ -2a \leq x
           \leq 2a,~~ a > 0  \label{e-mehta-dyson}.
    \eeq
$\rho$ is subject to positivity $\rho(x) \geq 0$ and normalization
$\int_{-2a}^{2a} \rho(y) dy = 1$ conditions. Since $S(x)$ is even,
$\rho$ must be even. It is convenient to introduce the generating
function of moments
    \beq
        F(z) = \int_{-2a}^{2a} {\rho(y) \over z-y} dy =
        \sum_{n=0}^\infty \fr{G_n}{z^{n+1}}.
    \eeq
(a) $F(z)$ is analytic on ${\mathbf{C}} \setminus [-2a,2a]$. (b)
$F(z) \sim \ov{z}$ as $|z| \to \infty$ which follows from
normalization. (c) $F(z)$ is real for real $z$ outside $[-2a,2a]$.
(d) When $z$ approaches $[-2a,2a]$, using (\ref{e-mehta-dyson}),
    \beq
    F(x \pm i\eps) = \half (4 x^3 - {2 x \over \nu + x^2}) \mp i \pi \rho(x)
    &\Rightarrow&
        \rho(x) = \ov{2\pi i} (F(x-i\eps) - F(x+i\eps))
    \eeq
(e) $F(z)$ is an odd function of $z$ since $\rho$ is even. By
analogy with the case of polynomial $S(x)$ we expect that if there
is an $F(z)$ satisfying these conditions, it is unique. Existence,
however, is not guaranteed. Indeed, we do not expect a 1-cut
solution for sufficiently small $\nu > 0$. In a region of validity
which is to be determined, the following ansatz for $F(z)$ is
consistent with the above requirements
    \beq
        F(z) = 2 z^3 - {z \over \nu + z^2} + R(z) \sqrt{z^2 - 4 a^2}.
    \eeq
$R(z)$ is a rational function to be chosen so that these conditions
are satisfied. In particular, we must pick $R(z)$ to cancel the
poles at $z = \pm i \sqrt{\nu}$ coming from ${2 z \over \nu + z^2}$.
So $R(z) = {P(z) \over (\nu + z^2)}$, where the polynomial $P(z)$
does not have zeros at $z = \pm i \sqrt{\nu}$. Moreover, to cancel
the linear and cubic terms in $S'(z)$ as $|z| \to \infty$, we need
to pick $P(z)$ to be a quartic even polynomial. Thus
    \beq
        F(z) = 2 z^3 - {z \over \nu + z^2} +
        {\a + \g z^2 + \eps z^4 \over (\nu + z^2)}
        \sqrt{z^2 - 4 a^2}.
    \eeq
We need to determine the parameters $a, \a, \g, \eps$. Though it may
appear that the third term is even in $z$, it is actually odd due to
the square root. As $z \to \infty$,
    \beq
    F(z) &\to& (2 + \eps) z^3 + \bigg[\g - \eps(2a^2 +
    \nu)\bigg] z \cr && +
    \bigg[ -1 + \a  - 2 a^4 \eps  - \g \nu  + \eps {\nu }^2 -
    2 a^2 (\g - \eps \nu) \bigg] \ov{z} + {\cal O}(\ov{z^3}).
    \eeq
Requiring $F(z) \sim \ov{z}$, fixes $\a, \g$ and $\eps$
    \beq
    \eps = - 2; ~~
    \g = - (2 \nu + 4 a^2); ~~
    \a = 2 - 12 a^4 - 4 \nu a^2.
        \label{e-alpha-gamma-eps}
    \eeq
To fix $a$ we must require analyticity of $F(z)$ as $z^2 \to - \nu$.
This is ensured if
    \beq
        -z + (\a + \g z^2 + \eps z^4) \sqrt{z^2 - 4 a^2}
    \eeq
vanishes as $z \to \pm i \sqrt{\nu}$. In other words,
    \beq
        \pm i \sqrt{\nu} + (\a - \g \nu + \eps\nu^2) \sqrt{-\nu - 4
        a^2}  = 0
    \eeq
or $-\nu  + {( -2 + 12 a^4) }^2 (4 a^2 + \nu)=0$, which is a quintic
equation for $s=a^2$,
    \beq
  576 s^{5} + 144 \nu s^4 - 192 s^3
            - 48 \nu s^2 + 16 s  + 3 \nu =0.
    \label{e-quintic-in-asq}
    \eeq
We are guaranteed at least one real solution for $a^2$. The
physically allowed solutions are those with $a^2 > 0$. For $\nu >
0$, we find that there are two solutions $a^2 > 0$ only the smaller
of which leads to $\rho(x)$ normalized to $1$ (the other has $\rho$
normalized to $3$). For this value of $a$, the density of
eigenvalues is
    \beq
        \rho(x) = - \bigg( \ov{\pi} \bigg) R(x) \sqrt{4 a^2 -  x^2}
        = - \bigg( \ov{\pi} \bigg) {(\a + \g x^2 + \eps x^4) \over (\nu + x^2)}
            \sqrt{4 a^2 - x^2}
    \label{e-one-cut-rho}
    \eeq
with $\a,\g,\eps$ given in (\ref{e-alpha-gamma-eps}).

\subsection{Moments}

\begin{figure}
\centerline{\epsfxsize=8.6truecm\epsfbox{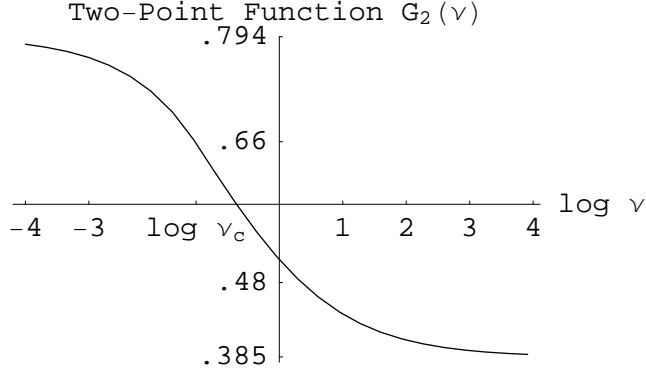}} \caption{Two-point
glue-ring expectation value $G_2(\n)$ plotted against $\log \n$. The
asymptotic behavior for heavy quarks or weak-coupling ($\n \to
\infty$, 1-cut phase) is separated from the behavior for gluons
strongly coupled to light quarks ($\n \to 0$, 2-cut phase) by a
third-order phase transition at $\n_c$. $G_2$ grows as we go from
gluons in the vacuum ($\n = \infty$) to those in a baryon strongly
coupled to light quarks (small $\n$) since $\rho(x)$ transforms from
uni-modal to bi-modal.} \label{f-G2}
\end{figure}

The analogue of glue-ring expectation values (\ref{e-def-of-Gn}) can
be obtained from the Laurent series for the moment generating
function $F(z)$
    \beq
        F(z) = 2 z^3 - {z \over \nu + z^2} +
        {\a + \g z^2 + \eps z^4 \over (\nu + z^2)}
        \sqrt{z^2 - 4 a^2} =
        \sum_{n=0}^\infty \fr{G_n}{z^{n+1}}
    \eeq
The odd moments vanish $G_{2n+1} = 0$ and the even ones are
    \beq
    G_2 &=& 40 a^6 + 12 \n a^4 - 4 a^2 - \n, \cr
    G_4 &=& 60 a^8 - 24 \n a^6 - 4(1+3 \n^2) a^4 + 4 \n a^2 +
        \n^2, ~~~~~~ {\rm etc}.
    \label{e-moments-one-cut}
    \eeq
where $a(\nu)= \sqrt{s}$ is the physical solution of
(\ref{e-quintic-in-asq}). $G_2(\n)$ and $G_4(\n)$ are plotted in
Fig.~\ref{f-G2} and \ref{f-G4}.

\subsection{Free energy}

\begin{figure}
\centerline{\epsfxsize=8.6truecm\epsfbox{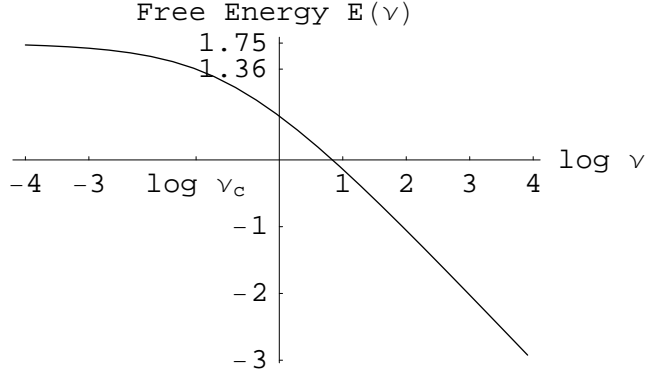}} \caption{Free
energy $E(\n)$ versus $\log \n$. The $2$-cut phase lies to the left
while the $1$-cut phase lies to the right of the critical point at
$\n_c$. For heavy quarks or weak-coupling ($\n \to \infty$), $E(\n)
\to -\log \n - .996$ as shown in Sec.~\ref{s-heavy-qk-limit} and
illustrated by the linear asymptotic behavior to the right in the
plot. In the chiral limit (Sec.~\ref{s-chiral-limit}), free energy
approaches a constant $E(0) \approx 1.751$.} \label{f-egy}
\end{figure}

Let us define the entropy \cite{entropy-var-ppl} as
    \beq
    \chi = {\cal P} \int \int \rho(x) \rho(y) \log{|x-y|} dx dy.
    \label{e-def-of-entropy}
    \eeq
The large-$N$ free energy is the Legendre transform of entropy, $E =
\int dx \rho(x) S(x) - \chi$. An expression for $E$ involving only
single integrals can be obtained using the Mehta-Dyson equation
    \beq
        \half S'(z) = {\cal P} \int {\rho(y) \over z-y} dy; ~~~ z
        \in {\rm supp}(\rho).
    \eeq
Integrating with respect to $z$ from $x_0$ to $x$, both of which lie
in the support of $\rho$,
    \beq
        \half \bigg[ S(x) - S(x_0) \bigg] = {\cal P} \int dy~
        \rho(y) ~ \bigg[\log{|x-y|} - \log{|x_0 -y|} \bigg].
    \eeq
Now multiply by $\rho(x)$ and integrate with respect to $x$,
    \beq
        \half \int dx ~\rho(x)~ \bigg[ S(x) - S(x_0) \bigg]
        = {\cal P} \int \int dx dy \rho(x) \rho(y)
        \bigg[\log{|x-y| - \log{|x_0-y|}} \bigg].
    \eeq
Thus
    \beq
        E &=& \half S(x_0) + {\cal P} \int dx ~\rho(x)~
            \bigg[\half S(x) - \log{|x-x_0|} \bigg] \cr
        E(\n) &=& \half S(x_0) + \half G_4 - {\cal P} \int_0^{2a} dx
            \rho(x) \log|(\n + x^2)(x^2 - x_0^2)|.
        \label{e-1-cut-free-energy}
    \eeq
$x_0$ is arbitrary provided it lies in the support of $\rho$. The
$x_0$ independence of $E$ follows from the Mehta-Dyson equation and
was also verified numerically. $x_0 = 0$ is most convenient for us,
so
    \beq
    E(\n) = - \half \log{\n} + \half G_4(\n) - {\cal P}
    \int_{0}^{2a} \rho(x) \log{(\n x^2 + x^4)} dx.
    \eeq
The new ingredient in free energy not determined by polynomial
moments $G_n$ is a sort of logarithmic moment. We could not evaluate
it in terms of known functions for the 1-cut $\rho(x)$
(\ref{e-one-cut-rho}), but the integration is easily performed
numerically and plotted in Fig.~\ref{f-egy}.

\subsection{Domain of validity of one-cut solution}

For which $\nu \geq 0$ is the 1-cut solution valid? So far we have
not imposed the $\rho \geq 0$ condition. The 1-cut solution
    \beq
    \rho(x) = -(1/\pi) \fr{P(x)}{(\nu + x^2)} \sqrt{4a^2 - x^2}; ~~~~~
    P(x) = \a + \g x^2 + \eps x^4
    \eeq
will break down if $\rho < 0$. The boundary of the region of
validity is given by $\nu$ for which $\rho(x) = 0$ for some $|x|
\leq 2a$. The most obvious way in which the 1-cut solution breaks
down is when $\rho(0) = 0$ and we transit to a 2-cut solution
without any support in an interval containing $x = 0$. $\rho(0) = -
\fr{2 a \a}{\pi \n}=0$ implies $\a = 0$, since $a > 0$. $\a = 0
\Rightarrow ~~~ 6 a^4 + 2 \n a^2 - 1 = 0$. Regarded as a quadratic
in $a^2$, the unique positive solution is
    \beq
    a^2_c = -\fr{\nu}{6} + \fr{\sqrt{\n^2 + 6}}{6} \geq 0 {\rm
    ~~~~for~~}   \nu > 0.
    \eeq
This is the critical value of $a(\n)$  at which the 1-cut solution
vanishes at the origin. Substituting $a_c^2$ into the quintic
(\ref{e-quintic-in-asq}) we find a condition on the transition point
$\n=\n_c$
    \beq
        \n \bigg[-27 - 8 \n^4 + 48 \n \sqrt{(6 + \n^2)}
        + 8 \n^2 (-9 + \n \sqrt{(6 + \n^2)}) \bigg] = 0.
    \eeq
Assuming $\nu > 0$, the only possibility is for the factor in
parentheses to vanish. Upon simplification it becomes a quadratic
equation $48 \n^4 + 368 \n^2 - 27 = 0$. The positive  solution is
    \beq
    \n_c = \sqrt{\fr{13 \sqrt{13} - 46}{12}} \approx 0.27
        \label{e-critical-point}.
    \eeq

Are there any other transition points, i.e. does $\rho(x)$ become
negative for any $0<|x|\leq 2a$? For example we find that there is
no value of $\n > 0$ for which $P(2a)=0$. Indeed, the condition
$P(2a)= 0$ implies that $30 a^4 + 6 \n a^2 -1 = 0$ or
    \beq
    a^2 = -\fr{\n}{10} + \fr{\sqrt{9 \n^2 + 30 }}{30}.
    \eeq
However, when this is substituted into the quintic
(\ref{e-quintic-in-asq}), the condition
    \beq
        4875 \n - 600 \n^3 + 72 \n^5 = \bigg(24 \n^4 - \fr{3200}{3}
            - 240 \n^2\bigg) \sqrt{30 + 9 \n^2}
    \eeq
has no solution for $\n > 0$. Based on the shape of $S(x)$ we expect
the 1-cut solution to be valid for all $\n>\n_c$, and we have
checked that this is indeed the case.

\section{Two-cut solution of quartic + log matrix model}
\label{s-2-cut-soln}

\begin{figure}
\centerline{\epsfxsize=8.6truecm\epsfbox{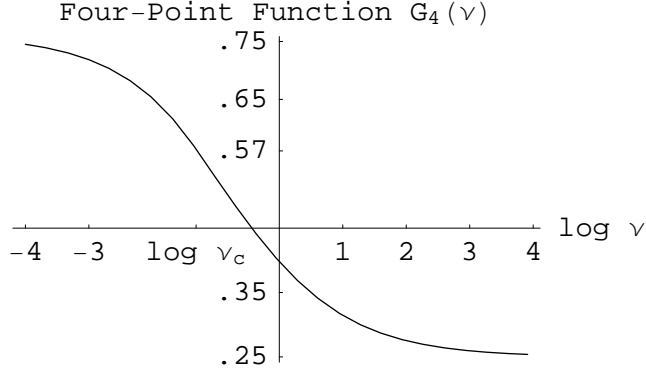}}
\caption{Four-point glue-ring expectation value $G_4(\n)$ versus
$\log \n$. The 1-cut solution for $\n \geq \n_c$ and 2-cut solution
for $\n \leq \n_c$ are not analytic continuations of each other
despite appearances. The asymptotic values at $\nu = 0, \infty$ and
at $\n_c$ shown on the vertical axis are obtained analytically in
Sec.~\ref{s-three-spl-cases}.} \label{f-G4}
\end{figure}

For small $\n$, $S(x) = x^4 - \log[\n + x^2]$ develops a repulsive
core near $x = 0$. For $\n<\n_c$, we expect the 1-cut solution of
the Mehta-Dyson equation
    \beq
    S'(x) = 4 x^3 - \fr{2x}{\nu + x^2} = 2 {\cal P}\int
        \fr{\rho(y)}{x-y} dy , ~~~~ x \in supp(\rho)
    \label{e-2-cut-mehta-dyson}
    \eeq
to make a transition to a 2-cut solution supported on $[-2a,-2b]
\cup [2b,2a]$. The generating function of moments $F(z) = \int
{\rho(y) \over z-y} dy$ enjoys the same properties $(a)$-$(e)$ as
before with $[-2a,2a]$ replaced with $[-2a,-2b] \cup [2b,2a]$ where
$a > b \geq 0$. An appropriate ansatz for $F(z)$ is
    \beq
        F(z) &=& \half S'(z) + R(z) \sqrt{(z^2-4a^2)}\sqrt{(z^2-4b^2)} \cr
        &=& 2 z^3 - {z \over (\nu + z^2)} +
        {(\beta z + \delta z^3) \over (\nu + z^2)} \sqrt{(z^2-4a^2)}\sqrt{(z^2-4b^2)}
    \eeq
$\beta, \delta, a, b$ have to be fixed using analyticity and
asymptotic behavior of $F(z)$. As $|z| \to \infty$,
    \beq
        F(z) &\to& \bigg(2 + \delta \bigg) z^3 + \bigg(\beta - \delta(2 a^2 +
        2 b^2 + \nu)\bigg) z \cr && + \bigg(-1 - 2\beta(a^2 + b^2) - 2 \delta(a^2 -
        b^2)^2 - \nu(\beta - 2 \delta (a^2 + b^2)) + \delta \nu^2 \bigg) {1 \over z} + {\cal O}({\ov{z^3}})
    \eeq
The requirement $F(z) \sim \ov{z} + {\cal O}(1/z^3)$ implies
    \beq
        \delta = -2 ; ~~~~
        \beta = -2(2a^2 + 2b^2 + \nu) ~~ {\rm~and~}~~
        6(a^4 + b^4) + 4a^2 b^2 + 2\n (a^2 + b^2) -1 = 0.
    \eeq
The condition that $F(z)$ be analytic at $z = \pm i \sqrt{\nu}$
implies $(\b - \n \d) \sqrt{(\n + 4 a^2)(\n + 4 b^2)} + 1 = 0$.
Substituting for $\b$ and $\d$, we are left with a pair of algebraic
equations for $a$ and $b$.
    \beq
        6(a^2 + b^2)^2 - 8a^2 b^2 + 2\n (a^2 + b^2) -1 &=& 0 \cr
        4(a^2 + b^2) \sqrt{16a^2b^2 + 4\n(a^2 + b^2) + \n^2} - 1
            &=&  0.
    \eeq
Let $s = a^2 + b^2$ and $p = a^2 b^2$ be the sum and product, then
    \beq
    6s^2 - 8p + 2 \n s -1 = 0  {\rm ~~~and ~~~}
    4s \sqrt{16p + 4\n s + \n^2} - 1= 0.
    \eeq
We can eliminate
    \beq
    p = \ov{4} \bigg[3 s^2 + \nu s - \ov{2} \bigg]
    \label{e-p-as-fn-of-s}
    \eeq
and get an algebraic equation for $s$, $4s \sqrt{12s^2 + 8 \n s +
(\n^2 - 2)} =1$, which has at most one positive solution $s$ for
$\n>0$. Squaring it we get a quartic equation
    \beq
    12 s^4 + 8 \nu s^3 + (\nu^2 - 2) s^2 -\ov{16} = 0.
    \label{e-alg-eqn-for-s}
    \eeq
This can be solved in lengthy but closed form and the unique
positive $s$ selected. From $s$, we get $p$ as well and
    \beq
        2a = \sqrt{2s + 2\sqrt{s^2 - 4p}},
         ~~~~ 2b =\sqrt{2s - 2\sqrt{s^2 - 4p}}.
    \eeq
The eigenvalue density, supported on $2b \leq |x|\leq 2a$ is
    \beq
    \rho(x) = -\fr{R(x)}{\pi} \sqrt{(4a^2 - x^2)(x^2 - 4 b^2)} =
        \fr{(2x^3+(4a^2+4b^2+2\n)x)}{\pi (\n + x^2)}
        \sqrt{(4a^2 - x^2)(x^2 - 4 b^2)}.
    \eeq
Using the Laurent expansion of $F(z)$, we get the moments
$G_{2n+1}=0$,
    \beq
    G_2 &=& 16s^3 - 64 ps - 8\n s^2 - 4\n^2 s + \n, \cr
    G_4 &=& 64 p^2 + 36 s^4 - 32 p s (5 s - \n) -
        8 s^3 \n - \n^2 + 8 s^2 \n^2 + 4s\n^3, ~~ \cdots
    \label{e-moments-two-cut}
    \eeq
$s$ and $p$ can be eliminated using the solution of the quartic
equation (\ref{e-alg-eqn-for-s}). The results are plotted in
Fig.~\ref{f-G2} and \ref{f-G4}.

\subsection{Free Energy}

The free energy may be expressed in terms of single integrals
    \beq
    E(\n) = \half S(x_0) + \half G_4
        - \half {\cal P} \int dx \rho(x) \log|(\n + x^2)(x^2 - x_0^2)|
    \label{e-2-cut-free-energy}
    \eeq
where $x_0$ is any point in the support of $\rho$. To obtain
(\ref{e-2-cut-free-energy}) we need to pay attention to the fact
that the Mehta-Dyson equation (\ref{e-2-cut-mehta-dyson}) is valid
in two disjoint intervals. Begin by integrating the Mehta-Dyson
equation with respect to $z$ from $x_+$ to $x$ where  $x_+,x \in
[2b,2a]$ to get\footnote{The principle value prescription is implied
where necessary and for brevity will not be explicitly indicated.}
    \beq
    \half[S(x)-S(x_+)] = \int dy \rho(y) [\log|x-y|-\log|x_+-y|]
    ~~~{\rm for}~~~~ x,x_+ \in [2b,2a]
    \eeq
Multiplying by $\rho(x)$ and integrating with respect to $x$ from
$2b$ to $2a$ and simplifying gives
    \beq
    \int_{2b}^{2a} dx \int dy \rho(x) \rho(y) \log|x-y| = \half
    \int_{2b}^{2a} dx \rho(x) S(x) - \fr{S(x_+)}{4} + \half
    \int dy \rho(y) \log|x_+-y|
    \eeq
for $x_+ \in [2b,2a]$. When the limits of integration are not
specified, the integral is over ${\rm supp}(\rho)$. Similarly, for
$x_- \in [-2a,-2b]$ we get
    \beq
    \int_{-2a}^{-2b} dx \int dy \rho(x) \rho(y) \log|x-y| = \half
    \int_{-2a}^{-2b} dx \rho(x) S(x) - \fr{S(x_-)}{4} + \half
    \int dy \rho(y) \log|x_- -y|.
    \eeq
Adding these two, we get for the entropy (\ref{e-def-of-entropy}),
    \beq
    \chi = \half \int dx \rho(x) S(x) - \fr{S(x_+)}{4} - \fr{S(x_-)}{4}
        + \half \int dy \rho(y) \log{|(x_+ - y)(x_- - y)|}
    \eeq
where $x_{\pm}$ are in the positive and negative part of
supp$(\rho)$. By choosing $x_- = - x_+$ we simplify matters using
the fact that $S(x)$ is an even function
    \beq
    \chi = \half \int \rho(x) S(x) dx - \half S(x_+) + \half \int
    dx~ \rho(x) \log|x^2 - x_+^2|.
    \eeq
Now observe that in this expression, the sign of $x_+$ does not
matter, so we can call $x_0 = x_+$ and pick it anywhere in
supp$(\rho)$. Recalling that $E = \int \rho(x) S(x) dx - \chi$ we
get the advertised expression (\ref{e-2-cut-free-energy}) for the
2-cut free energy. Though the complexity in evaluating $E(\n)$ has
been reduced, we have not been able to find the above logarithmic
moment in terms of known functions. The numerically evaluated free
energy $E(\n)$ is plotted in Fig.~\ref{f-egy}.

\subsection{Phase transition to one-cut solution}

We expect the $2$-cut solution to make a transition to the $1$-cut
solution when the intervals $[-2a,-2b], [2b,2a]$ merge, i.e. $b=0$,
which implies $p = 0$ and $s = a^2$. Inserting in
(\ref{e-p-as-fn-of-s}) gives a quadratic equation $3s_c^2 + \n s_c -
\half = 0$ whose positive solution $s_c = a_c^2 = -\fr{\n}{6} +
\fr{\sqrt{\nu^2 + 6}}{6}$ is the same as the value of $a^2$ at which
the 1-cut solution breaks down. When $s_c$ is substituted in
(\ref{e-alg-eqn-for-s}), we get the same phase transition point
$\n=\n_c$ as before (\ref{e-critical-point}). The $2$-cut solution
for $\n < \n_c$ takes over when the 1-cut solution breaks down.

\section{Special cases: weak-coupling, critical point and chiral limit}
\label{s-three-spl-cases}

\subsection{Heavy quark or weak-coupling limit $\n \to \infty$}
\label{s-heavy-qk-limit}

In our toy-model, when the quarks are very heavy or the coupling
constant is small, the self-interactions of the gluons dominates
their interactions with the quarks in the baryon. This is the limit
$\n \to \infty$ where the action $S(A) \to \tr A^4$ up to an
additive constant. It is as if the gluons do not feel the presence
of the baryon and we return to calculating {\em vacuum} correlations
of glue-ring observables. This limit lies in the deep end of the
1-cut phase, where calculations simplify. The quintic equation
(\ref{e-quintic-in-asq}) for the limits of $supp(\rho)$ reduces to a
quadratic equation $48 s^4 - 16 s^2 + 1 = 0$ whose physical solution
is $s = \sqrt{3}/6$. The limiting eigenvalue density is
    \beq
        \rho(x,\n \to \infty) = \ov{\pi} (4s+2x^2) \sqrt{4s-x^2},
        ~~~~ |x| \leq 2 \sqrt{s},
    \eeq
The odd moments vanish while the even moments and free energy
(\ref{e-1-cut-free-energy}) are
    \beq
    G_{2n} = \fr{2^{n+1} (n+1) \G(n+\half)}{3^{n/2} \sqrt{\pi}
    \G(n+3)}; && G_2 = \fr{2\sqrt{3}}{9} \approx .385, ~~~ G_4 = \ov{4},
    \cdots \cr
    E = -\log \n + \half G_4 - \half \int dx~ \rho(x) \log x^2 &=& - \log \n + \ov{8}(3+
    \log{144}) \approx - \log \n + .996 .
    \eeq
These limiting values are seen to agree with the numerically
obtained behavior of $E(\n)$, $G_2(\n)$ and $G_4(\n)$ plotted in the
Fig.~\ref{f-egy}, \ref{f-G2} and \ref{f-G4} for a wide range of
values of $\n$.

\subsection{Neighborhood of the phase transition}
\label{s-nbd-of-ph-trans}

\begin{figure}
\centerline{\epsfxsize=8.6truecm\epsfbox{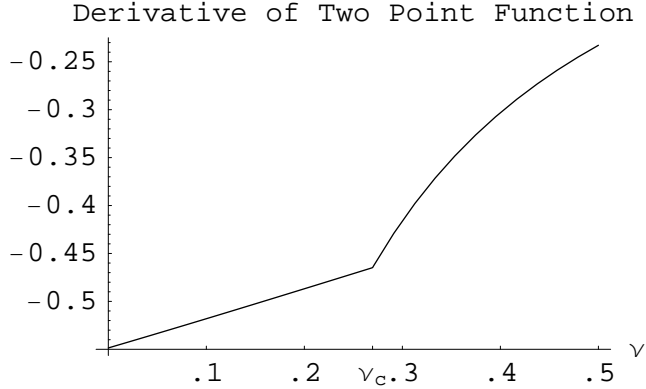}}
\caption{Derivative of the two-point correlation $G_2(\n)$ has a
kink at $\n_c=0.27$ indicating that its second derivative is
discontinuous.} \label{f-dG2}
\end{figure}

\begin{figure}
\centerline{\epsfxsize=8.6truecm\epsfbox{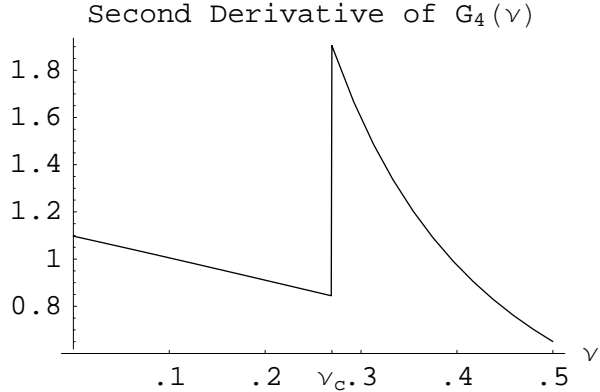}} \caption{Jump
discontinuity in the second derivative of $G_4(\n)$ at the critical
point $\n_c$.} \label{f-ddG4}
\end{figure}

\begin{figure}
\centerline{\epsfxsize=8.6truecm\epsfbox{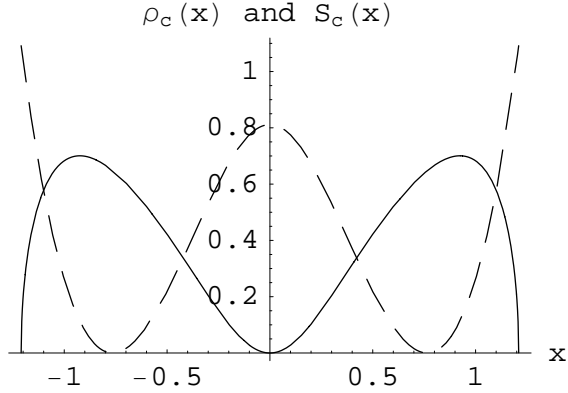}}
\caption{Eigenvalue density at the critical point $\rho_c(x)$, the
continuous curve. The dashed curve is the critical action
$S(x,\n_c)$ plus a constant chosen to make it vanish at its minima.
While $\rho$ is peaked around the minima of $S$, the spread allows
it to maximize entropy $\chi$ (\ref{e-def-of-entropy}).}
\label{f-rhocrit}
\end{figure}

At $\n_c$, the 1-cut solution, valid for high quark mass $m$ or
weak-coupling $\a$ makes a phase transition to a 2-cut solution,
which is valid for low quark mass or strong-coupling ($\n \leq
\n_c$). In the immediate vicinity of the phase transition,
observables are more easily evaluated than generically. We find that
the eigenvalue density is continuous across the transition, the
critical eigenvalue density is (see Fig.~\ref{f-rhocrit})
    \beq
    \rho_c(x) = \fr{\{ (7\sqrt{13}-22)\n_c x^2 + (13\sqrt{13} -46) x^4 \}}
        {18\pi \n_c^2 (\n_c + x^2)} \sqrt{(8\sqrt{13}+20)\n_c - 9x^2}
    \eeq
and is supported on $[-2a_c,2a_c]$ where $a_c^2 = \sqrt{(
\fr{\sqrt{13}-2}{12})}$, $a_c \approx .605$. Thus, all moments and
the free energy are also continuous.
The next question is whether their derivatives are discontinuous
across the transition. We find that the second derivatives of $G_2$
and $G_4$ have jump discontinuities across the phase transition. We
have calculated this behavior analytically. To find the derivatives
of the correlations
(\ref{e-moments-one-cut}),(\ref{e-moments-two-cut}) at $\n_c$ we
need the values and derivatives of $a, s$ and $p$ at $\n_c$. The
critical values $\n_c = \fr{\sqrt{13\sqrt{13}-46}}{12},
s_c=a_c^2,p_c=0$ have already been determined without much trouble.
For their derivatives, it is again not necessary to solve the
quartic (\ref{e-alg-eqn-for-s}) and quintic (\ref{e-quintic-in-asq})
equations, but suffices to solve linear equations obtained by
differentiating these at $\n_c$. For example, suppose we want
$G_2^\pr(\n_c^-)$, which corresponds to the approach from the 2-cut
phase. From (\ref{e-moments-two-cut})
    \beq
    G_2^\pr(\n) &=& 48 s^2 s' - 64 p s' - 64 s p' -4 \n^2 s' - 8 s \n
    -8s^2 -16 ss' \n + 1
    \eeq
Differentiating (\ref{e-alg-eqn-for-s}) and solving the linear
equation for $s^\pr(\n)$ gives
    \beq
    s^\pr(\n) = \fr{-\n s - 4s^2}{-2+\n^2+12 \n s + 24s^2}.
    \eeq
$p'(\n)$ is determined similarly using (\ref{e-p-as-fn-of-s}) and
evaluated at $\n_c$. In this manner we get
    \beq
    G_2(\n_c^\pm)= \fr{(13\sqrt{13} + 19)}{27} \n_c \approx .66;
        && G_2^\pr(\n_c^\pm) = \fr{\sqrt{13}-5}{3} \approx -.46 \cr
    G_2^{\pr \pr}(\n_c^+) = \fr{4}{9}\sqrt{\fr{(205\sqrt{13}
        -122)}{39}} \approx 1.77; &&
    G_2^{\pr \pr}(\n_c^-) = \fr{8}{51}
        \sqrt{\fr{(1669\sqrt{13}-5858)}{39}} \approx 0.32
    \eeq
and
    \beq
    G_4(\n_c) = \fr{(26\sqrt{13}+119)}{27} \n_c^2 \approx .57; && G_4^\pr(\n_c) \approx .53 \cr
    G_4^{\pr \pr}(\n_c^-) = \fr{(4706-1150\sqrt{13})}{663} \approx 0.844;
    &&
    G_4^{\pr \pr}(\n_c^+) = \fr{(2522 - 514\sqrt{13})}{351} \approx 1.905
    \eeq
These agree with the numerically determined correlations plotted in
Fig.~\ref{f-dG2} and Fig.~\ref{f-ddG4}.

\begin{figure}
\centerline{\epsfxsize=8.6truecm\epsfbox{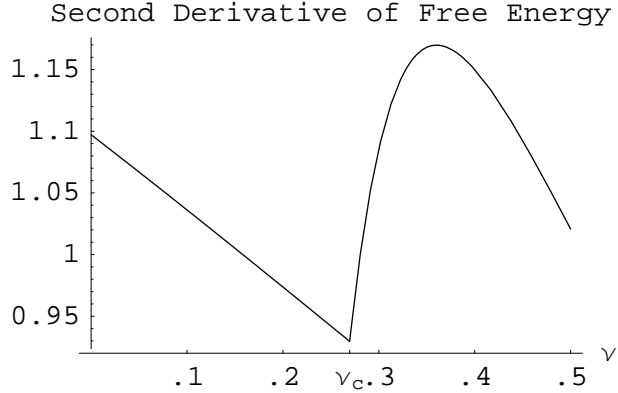}}
\caption{$E^{\pr \pr}(\n)$ versus $\n$. The $3^{\rm rd}$ derivative
of free energy is discontinuous at $\n_c=0.27$.} \label{f-ddegy}
\end{figure}

The free energy $E(\n)$ and its first two derivatives are continuous
at $\n_c$. $E^{\pr \pr \pr}(\n)$ has a jump discontinuity at $\n_c$.
For example, to calculate $E^\pr(\n_c^+)$, we differentiate the
integral representation for 1-cut free energy
(\ref{e-1-cut-free-energy})
    \beq
    E(\n) &=& -\half \log \n + \half G_4 - \int_0^{2a} dx~
        \rho_\n(x) \log{(\n x^2 + x^4)} \cr
    \Rightarrow ~~ E^{\pr}(\n) &=& -\ov{2\n} + \half G_4^\pr(\n) -
        \int_0^{2a} dx \bigg\{ \fr{\rho(x)}{\n + x^2}
        + \dd{\rho}{\n} \log{(\n x^2 + x^4)} \bigg\}.
    \eeq
$\dd{\rho}{\n}$ can be got from the explicit formula
(\ref{e-one-cut-rho}). We omitted the term involving the derivative
of the upper limit of integration because $\rho(2a) =0$. Evaluating
at $\n_c$, using the above result for $G_4^\pr(\n)$ and doing the
integral numerically gives us $E^\pr(\n_c^+)$. Proceeding along
these lines we get $E(\n_c) \approx 1.36, E^\pr(\n_c) \approx -1.32,
E^{\pr \pr}(\n_c) \approx .93, E^{\pr \pr \pr}(\n_c^-) \approx -.7,
E^{\pr \pr \pr}(\n_c^+) \approx 7.05$ This is illustrated in
Fig.~\ref{f-ddegy}. We conclude that the phase transition is of
third-order.

\subsection{Chiral limit or strong-coupling limit $\n \to 0$}
\label{s-chiral-limit}

In the limit of massless quarks or strong-coupling, $\n \to 0^+$,
which lies in the 2-cut phase. The action becomes $S(A)=\tr[A^4 -
\log(A^2)]$ and the gluons maximally feel the presence of quarks in
the baryon. The solution again simplifies. The quartic equation for
$s$ (\ref{e-alg-eqn-for-s}) becomes a quadratic with solution
$s=\half \sqrt{\fr{2+\sqrt{7}}{6}}$ and $p = \fr{\sqrt{7}-2}{32}$.
From this we get
    \beq
        \rho(x)= \fr{(2x^2 + 4s)}{\pi x} \sqrt{4 s x^2 -x^4 -16 p}
    \eeq
and moments $G_2 = \fr{(4-\sqrt{7})}{3} \sqrt{\fr{4+2\sqrt{7}}{3}}
\approx .794$ and $G_4 = \fr{3}{4}$ which are shown in
Fig.~\ref{f-G2} and \ref{f-G4}. The free energy
(\ref{e-2-cut-free-energy}) has a finite limit $E(\n=0) \approx
1.751$ as illustrated in Fig.~\ref{f-egy}.

\section{Discussion}
\label{s-discussion}

We have studied a very simple matrix model for glue-ring
correlations in a baryon in the limit of many colors. It was
obtained as a caricature of the dimensional reduction of QCD to
$1+1$ dimensions. Our main finding is that there is a third-order
phase transition that separates a phase where gluons are weakly
coupled to heavy quarks ($\n \geq \n_c$) from one where the quarks
are light and strongly coupled to gluons ($0 \leq \n \leq \n_c$).
$\n$, a dimensionless ratio of quark mass to coupling constant is
the only parameter of the model. While for $\n \geq \n_c$ we have a
1-cut solution of the matrix model, for $\n \leq \n_c$ we have a
2-cut solution. The gauge-invariant observables (glue-ring
expectation values) are described in these two phases by two
different analytic functions of $\n$ that disagree beyond their
first derivatives at $\n_c$ (See Fig.~\ref{f-G2}, \ref{f-G4},
\ref{f-dG2}, \ref{f-ddG4}). Moreover, the case of gluons in the
vacuum i.e. where the baryon is absent, corresponds to $\n \to
\infty$, which is deep inside the 1-cut phase. The physically
interesting value of $\n$ most likely is small and lies in the 2-cut
phase, since current quarks in the proton are very light compared to
$\La_{\rm QCD}$. Thus, the vacuum correlations of gluons are likely
to be separated by a phase transition from those in a baryon state.
Moreover, from Fig.~\ref{f-G2} and \ref{f-G4} we see that gluon
correlations are enhanced inside the baryon compared to their values
in the vacuum $(\n = \infty)$. This reflects the growth of moments
of the distribution of eigenvalues, as they make a transition from
being clustered about the origin to being supported on a pair of
intervals excluding the origin. Though this is very far from
explaining why about half the proton's momentum is contributed by
gluons, it does indicate that the qualitative features of gluon
correlations in the vacuum can be quite different from those in a
baryon state. Even if we could use a weak-coupling expansion to
describe a bound-state like a baryon, the phase transition would
invalidate analytic continuation to the physically relevant baryon
containing gluons strongly coupled to light quarks.

The wider applicability of our results is called into question by
our approximations and truncations. The sharp phase transition is an
artifact of $N =\infty$. For finite $N$, the matrix model has
finitely many degrees of freedom and cannot display non-analytic
behavior. Nevertheless, this is probably the most benign of our
approximations. The finite $N$ theory should display qualitative
differences between the two regimes. Absence of space-time
derivatives and non-local interactions due to longitudinal gluons
are the more significant shortcomings of our model. This toy-model
has given us a cartoon of how the theory may behave as $\n$ is
varied. A hamiltonian approach may have a better chance at shedding
light on the matrix field theory or matrix quantum mechanics version
of this problem. We hope the proposal of treating gluons in a `fixed
baryon background' $| \Psi \ket$ (\ref{e-baryon-state}), along with
other new ideas will help simplify the matrix field theory in order
to better understand the emergent bound-state structure of gluons in
a nucleon.

Some questions for future work are collected here. (i) Can we use a
variational or other approximation method to understand this phase
transition? Such an approach has a better chance of generalizing to
multi-matrix models. (ii) Besides glue-ring correlations, we are
also interested in open string correlations in a baryon state. Can
we get a zero-dimensional toy-model for these as well? (iii) Can we
shed any light on the multi-matrix model that arises when we do {\em
not} assume the transverse gluon field to be equal at the positions
of the quarks? (iv) What is the relation between the gluon
distribution function extracted from experimental data and the
gauge-invariant glue-ring variables that would come from solving the
matrix field theory? One suspects that the gluon distribution is
essentially determined by a two-point glue-ring expectation value.
What is the simplest truncated form of the matrix field theory where
such a gluon distribution function can be estimated? (v) In
\cite{entropy-var-ppl}, we obtained algebraic and probabilistic
characterizations of the entropy $\chi$ whose Legendre transform is
the free energy of matrix models. However, the formula for $\chi$ in
generic multi-matrix models is quite complicated. Is there any
multi-matrix generalization of the trick (\ref{e-1-cut-free-energy})
of using the Mehta-Dyson `equation of motion' to reduce double
integrals to single integrals? (vi) Can we find a variational
principle that determines the closed and open string observables of
our matrix field theory or some finite dimensional truncation
thereof? Such a variational principle for closed string observables
in multi-matrix models was found in \cite{entropy-var-ppl}.

\section*{Acknowledgements}

The author thanks S. G. Rajeev and G. 't Hooft for discussions and
acknowledges support of the European Union in the form of a Marie
Curie Fellowship.



\end{document}